\documentclass[aps,prl,reprint,superscriptaddress,floatfix,nofootinbib]{revtex4-1}
\usepackage{hyperref}
\hypersetup{colorlinks=true, citecolor=blue, urlcolor=blue, linkcolor=blue}
\usepackage{graphicx}
\usepackage[normalem]{ulem}
\usepackage{amsmath}
\usepackage{color}
\usepackage{textcomp}
\usepackage{gensymb}
\usepackage[top=1in, bottom=1in, left=0.5in, right=0.5in]{geometry}
\usepackage{svg}
\usepackage{pgf}
\usepackage{multirow}

\usepackage{booktabs}

\graphicspath{{./}}

\begin{document}
\title{Charge-noise induced dephasing in silicon hole-spin qubits}
\author{Ognjen Malkoc}
\affiliation{RIKEN Center for Emergent Matter Science, Wako-shi, Saitama 351-0198, Japan}
\author{Peter Stano}
\affiliation{RIKEN Center for Emergent Matter Science, Wako-shi, Saitama 351-0198, Japan}
\author{Daniel Loss}
\affiliation{RIKEN Center for Emergent Matter Science, Wako-shi, Saitama 351-0198, Japan}
\affiliation{Department of Physics, University of Basel, Klingelbergstrasse 82, CH-4056 Basel, Switzerland}
\begin{abstract}
We investigate theoretically charge-noise induced spin dephasing of a hole confined in a quasi-two-dimensional silicon quantum dot. Central to our treatment is accounting for higher-order corrections to the Luttinger Hamiltonian. Using experimentally reported parameters, we find that the new terms give rise to sweet-spots for the hole-spin dephasing, which are sensitive to device details: dot size and asymmetry, growth direction, and applied magnetic and electric fields. Furthermore, we estimate that the dephasing time at the sweet-spots is boosted by several orders of magnitude,
up to order of milliseconds. 
\end{abstract}

\maketitle

{\it Introduction.} Silicon is promising for realizing scalable qubits using quantum-dot electrons to store and process quantum information 
\cite{loss_quantum_1998,chatterjee_2021,stano_2021,burkard_2021}.
The recent attention to silicon stems from compatibility with industrial fabrication \cite{thompson_uniaxial-process-induced_2006,gonzalez_2021,zwerver_2021,vinet_2021} 
and low noise from nuclear spins. The latter effect, so far a major obstacle for spin qubits in GaAs, can be further suppressed using holes instead of electrons 
\cite{bulaev_spin_2005,heiss_2007,chekhovich_direct_2011,prechtel_decoupling_2016}.
Holes also offer stronger spin-orbit coupling 
\cite{bogan_landau-zener-stuckelberg-majorana_2018,venitucci_simple_2019,camenzind2021,bosco_hole_2020}, essential for electric spin control without micromagnets or on-chip ESR lines \cite{maurand_cmos_2016}.
Taken together, the reduced susceptibility to nuclear noise \cite{boscoprl2021}, 
the absence of valley degeneracy, and fully-electric control make holes in silicon a very attractive platform for scalable spin qubits.

In a quasi-two-dimensional (planar) quantum dot,with the strongest confinement along the growth direction, the confinement splits the fourfold degeneracy at the $\Gamma$ point into light and heavy holes, offering a resilient spin qubit residing in the heavy hole subspace \cite{bulaev_spin_2005, bulaev_electric_2007,martin_two-dimensional_1990}. Spin blockade detection \cite{li_pauli_2015, bohuslavskyi_pauli_2016, yamaoka_charge_2017, wang_anisotropic_2016}, control over the charge state down to a single hole \cite{liles_spin_2018, sousa_de_almeida_ambipolar_2020}, fabrication of arrays \cite{hendrickx_four-qubit_2020, lawrie_quantum_2020, van_riggelen_two-dimensional_2020}, and demonstration of single \cite{hofmann_assessing_2019, hendrickx_single-hole_2019} and two-qubit operations \cite{hendrickx_fast_2020} are among recent experimental achievements with planar dots.
In contrast, the strong confinement-induced spin-orbital mixing in a nanowire geometry \cite{hu_hole_2011, pribiag_electrical_2013, gao_sitecontrolled_2020} gives large and tunable spin-orbit interaction
 \cite{kloeffel_strong_2011,froning_2021,kloeffel_direct_2018,bosco_benito_2021} and fast spin manipulation \cite{froning_ultrafast_2021, wang_ultrafast_2020}. 

The strong sensitivity to the electric field is a generic feature of hole-spin qubits. 
Among its most direct manifestations, the electrical response of the g-factor has been reported for various designs \cite{andlauer_electrically_2009, katsaros_hybrid_2010, klotz_observation_2010,ares_nature_2013, bennett_voltage_2013, pribiag_electrical_2013, prechtel_electrically_2015, brauns_electric-field_2016, voisin_electrical_2016, de_vries_spinorbit_2018, crippa_electrical_2018}.
While it offers increased electrical tunability, it also implies a higher susceptibility to electrical noise.
With nuclear noise negligible, charge noise becomes the primary concern for qubit coherence \cite{houel_high_2014, maurand_cmos_2016, yoneda_quantum-dot_2017}. Aiming at long coherence time, the most favorable scenario seems to be a single hole in an isolated planar quantum dot. Assessing this ultimate limit on the hole-spin coherence is our main objective.

We find that in planar dots the spin-electric coupling is dominated by higher-order (non-quadratic in momentum) terms, which are not contained in the often used and well-known Luttinger Hamiltonian \cite{luttinger_quantum_1956}. This finding is among our main results.

\begin{figure}
	\centering
	\centering
	\includegraphics[width=0.85\linewidth]{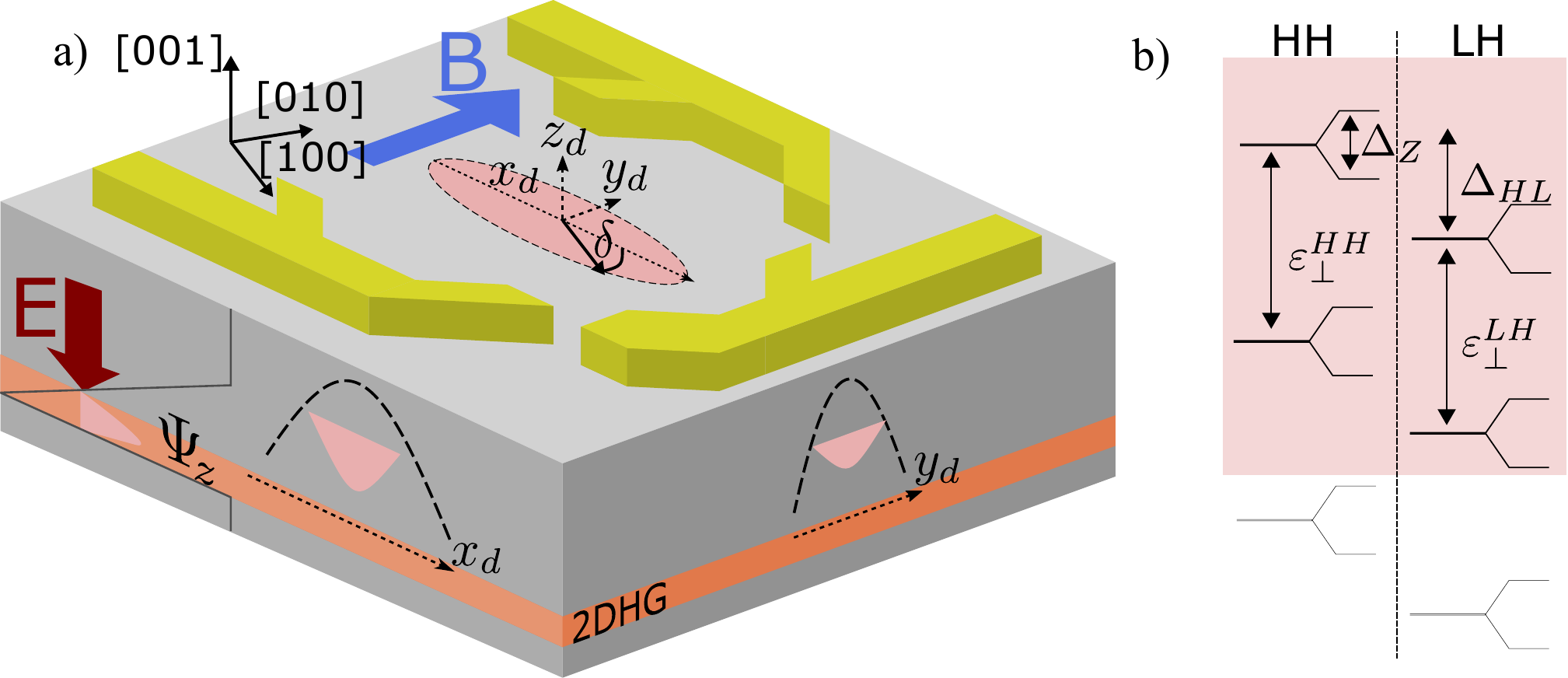}
\caption{ a) A schematic of a lateral quantum dot hosting a spin qubit. A hole from the two-dimensional-hole-gas (2DHG) is trapped by the confinement potential, that has a major axis making an angle $\delta$ with the crystallographic [100] axis.  b) The basis states in the calculation, showing the heavy hole and light hole subbands of the unperturbed Hamiltonian. The red area shows the states used in the perturbation series of Eq.~\eqref{eq:perturbation}.
	}
	\label{Fig:DeviceAndBandSchematic}
\end{figure}

While the conduction band non-parabolicity has been studied in zinc-blende crystals in detail \cite{ogg_conduction-band_1966,rossler_nonparabolicity_1984}, including its effects on the g-factor \cite{stano_g_2018}, 
the valence band requires a separate treatment.
To this end, we derive the corrections to the Luttinger Hamiltonian up to the fourth order in momentum 
and up to linear in the electric field. 
Even though one can generate these terms by symmetry analysis, for example using the tables in Ref.~\cite{winkler_spin-orbit_2003}, we are not aware of their prefactors being known. To evaluate the spin-orbit effects reliably, however, these prefactors are necessary.
We calculate them within the 14-band $k\cdot p$ model \cite{pfeffer_five-level_1996}, using up to the fifth order of the L\"owdin perturbation theory.
The resulting effective model is valid for any materials with diamond crystal structures, such as Si and Ge, but we will focus on the former material.
By focusing exclusively on Si devices, we use the Si band structure parameters and identify a minimal set, 
given in Eq.~\eqref{eq:NewTerms}, 
which, 
together with Eq.~\eqref{Eq:Luttinger}, 
describes the essential correction terms to the valence band in the bulk 
 for describing spin dephasing.

We obtain the spin-qubit Hamiltonian by projecting the valence band Hamiltonian onto the lowest orbital state, defined by the three-dimensional confinement potential.
Below, we will use  second order perturbation theory to include the effects of higher orbital states.
From the qubit Hamiltonian
we evaluate two quantities of interest: the $g$-tensor and the dephasing rate. 
Our main result in this part is twofold. First, the typical dephasing time is found to be on the order of tens of microseconds. 
This value is then the ultimate upper limit in any design  with holes in silicon gated dots with an in-plane magnetic field. With other than planar dots one would expect the dephasing time to be much smaller 

Second, we find pronounced sweet spots, where the dephasing time is boosted up to milliseconds. Their position in parameter space is sensitive to all system parameters. The suggestion to search for experimentally robust sweet spots is the main practical implication of our work.

{\it Valence band corrections.} 
All symmetry-allowed terms in the valence band of silicon up to quadratic in the kinetic momentum $\hbar \mathbf{k}$ are contained in the Luttinger Hamiltonian \cite{luttinger_quantum_1956}, 
\begin{equation}
	\begin{aligned}
		\!\!  H_{L}& =  \frac{\hbar^2}{2m_0}\!\left[\! -\!\!\left(\! \gamma_1\!\! +\!\! \frac{5 \gamma_2}{2} \! \right)\! k^2\!\! + \!2\gamma_2 (k_X^2 J_X^2\!\! +\! k_Y^2J_Y^2\!\!+\! k_Z^2 J_Z^2) \right. \\
	\!\!	+ 4\gamma_3 &(k_{XY}J_{XY}\!\!+\! k_{YZ}J_{YZ}\!\! +\! k_{ZX} J_{ZX} \! )\! \Big]   \!  \!     - \!  2  \mu_B( \kappa \mathbf{J}\! \!+\! q \mathbf{J}_3) \!\cdot\!\mathbf{B}.
	\end{aligned}
\label{Eq:Luttinger}
\end{equation}
Here, the components of the vectors $\mathbf{J} = (J_X, J_Y, J_Z) $ and $\mathbf{J}_3 =( J_X^3, J_Y^3, J_Z^3)$, are the spin 3/2 operators, $m_0$ is the free-electron mass,
$A_{ij}=A_i A_j+A_j A_i$ 
is the anticommutator, and the coefficients $\gamma_{1},\gamma_{2},\gamma_{3}, \kappa$, and $q$ are the Luttinger parameters \footnote{We use the experimentally measured values $\gamma_1 = 4.285, \gamma_2 = 0.339, \gamma_3 =1.446 , \kappa = -0.42, \text{ and }  q = 0$, instead of calculating them within the 14-band model, since the latter procedure underestimates these parameters.}. Also, $X$, $Y$, and $Z$ denote the [001], [010], and [001] crystallographic axis, respectively.
Finally, due to the relation $\mathbf{k}\times \mathbf{k} = -ie\mathbf{B}/\hbar$, the magnetic $\mathbf{B}$-field  components are counted as quadratic in momentum.

Using symmetry analysis, we derive corrections to Eq.~\eqref{Eq:Luttinger} up to the fourth order in momentum and up to linear in the electric field. For these two groups we get fifteen and twelve terms, respectively. We evaluate their prefactors using the 14-band $k\cdot p$ model, in the fourth and fifth order of the L\"owdin perturbation theory \cite{lowdin_note_1951},  respectively. With the full list including formulas for prefactors given elsewhere, we restrict here ourselves to an excerpt. Namely, after analyzing each of the 27 terms as outlined below, we identify the terms  contributing dominantly to the spin dephasing for devices grown along [001] \footnote{The set of terms depends on the growth orientation of the device. An analogous set of terms which dominate devices grown along [111] in Eq.~\ref{Eq:NewTerms111} of the Supplemental Material.}. There are magnetic-field generated terms,
\begin{subequations}
\label{eq:NewTerms}
\begin{equation}
	\begin{aligned}
		H_{41}= & \mu_B (\kappa_{41} \mathbf{J}  + 	q_{41} \mathbf{J}_3)\cdot \mathbf{B} (k_X^2 + k_Y^2 + k_Z^2),                            \\
		H_{42}= & \mu_B(\kappa_{42} \mathbf{J} + q_{42}\mathbf{J}_3) \cdot  (B_X  k_X^2, B_Y  k_Y^2, B_Z  k_Z^2), \\
		H_{43}= & \mu_B(\kappa_{43} \mathbf{J} + q_{43}\mathbf{J}_3) \cdot  (k_X (k_Y B_Y + k_Z B_Z), \mathrm{c.p.}),                  \\
		H_{53}= & \mu_B \Gamma_{53}\mathbf{J}_{53} \cdot  (B_X(k_Y^2-k_Z^2), \mathrm{c.p.}),                  \\
	\end{aligned}
	\label{eq:NewTermsFirst}
\end{equation}
and the band-warping terms, 
\begin{equation}
	\begin{aligned}
	H_{12}= &\Gamma_{12} (\{k_X,k_Y\}^2 +\{k_Y,k_Z\}^2  +\{k_Z,k_X\}^2 ),                            \\
	H_{32}= & \Gamma_{32} \mathbf{J}_{32} \cdot (2k_X^2 k_Y^2 \!-\! (k_Y^2+ k_X^2)k_Z^2,  (k_Y^2  - k_X^2)k_Z^2).
	\end{aligned}
	\label{eq:NewTermsSecond}
\end{equation}
\end{subequations}
In these equations, $\mathbf{J}_{53} = (\{J_x,J_y^2-J_z^2\},\{J_y,J_z^2-J_x^2\},\{J_z,J_x^2-J_y^2\})$, $\mathbf{J}_{32} = (J_Z^2 - \mathbf{J} \cdot \mathbf{J}/3, J_X^2 - J_Y^2)$, and $\mathrm{c.p.}$ means cyclic permutation. 
The values of prefactors are given in Tab.~\ref{ParameterTable} of the Supplemental Material \cite{SM}. The figures in the following sections are plotted using all 27 corrections, together denoted as $\delta H_L$. In the Supplementary Material (SM)~\cite{SM}, we show analogous figures produced with Eq.~\eqref{eq:NewTerms} instead. One finds that the latter is an excellent approximation.

\newcommand{\valueBandP}{ 0.871  }
\newcommand{\valueBandQ}{0.750  }
\newcommand{\valueEZero}{ 4.185 }
\newcommand{\valueEZeroPrime}{ 3.4   }
\newcommand{\valueDeltaZero}{ 0.044    }
\newcommand{\valueDeltaZeroPrime}{ 0 }
\newcommand{\valueKappa}{-0.84}
\newcommand{\valueQ}{  0 }
\newcommand{\valueKappaFourOne}{-0.15}
\newcommand{\valueQFourOne}{0.71}
\newcommand{\valueKappaFourTwo}{79.81}
\newcommand{\valueQFourTwo}{-32.46}
\newcommand{\valueKappaFourThree}{-78.46}
\newcommand{\valueQFourThree}{31.05}
\newcommand{\valueGammaFiveOne}{-1804.6}
\newcommand{\valueGammaFiveThree}{-0.71}
\newcommand{\GammaOneTwo}{939.59}
\newcommand{\GammaThreeTwo}{456.99}

{\it Effective hole-spin qubit Hamiltonian.} We consider a hole confined in a device shown in Fig.~\ref{Fig:DeviceAndBandSchematic}(a).
Since the spin-orbit interaction is anisotropic, as can be seen already in $H_{L}$, the splitting gap depends on the growth direction: \cite{takahashi_experimental_2011}
changing it from [001] to [111], in our model the gap increases from $2.1\text{ meV}$ to $29.7\text{ meV}$ for a triangular vertical confinement with 10 V/$\mu$m electric field. 
A similar splitting arises from strain 

\cite{hendrickx_gate-controlled_2018, sun_physics_2007, hardy_single_2019}. For instance, a strain of $0.5\%$
in Si would give a splitting of order $10$ meV \footnote{The estimation is for bi-axial strain in the dot-plane, where $\Delta_{strain}=|b|(1+c_{11}/c_{12})\varepsilon_{\parallel}$, where $\varepsilon_{\parallel}$ is the in-plane strain. The deformation potential $b=2.2$ and elastic coefficients $c_{ij}$ are taken from Ref.~\cite{winkler_spin-orbit_2003}.}. The principal difference is which spin is the ground state: with strain, depending on its type, it can be either the heavy or the light hole 

\cite{huo_light-hole_2014,lodari_light_2019}. On the other hand, in the planar geometry the heavy hole ground state always results from the vertical confinement alone \cite{fischetti_six-band_2003}.
We assume that the heavy hole subspace is the ground state and the qubit is defined therein, as a configuration most resilient to charge noise. 

We adopt standard choices to describe the quantum dot confinement: a triangular potential for the vertical part and an anisotropic harmonic for the in-plane part,
\begin{equation}
	V_{xy}   = \frac{m^{3/2}_{xy}}{2 \hbar^2} (\varepsilon_x^2 x^2 + \varepsilon_y^2 y^2 ), \quad
	V_{z}  =  \begin{cases}
		e E_z z & \text{for $z>0$}    \\
		V_0       & \text{for $z\leq0$}
	\end{cases}	.
\end{equation}
Here, $V_0$ is the heterostructure band offset, $m^{3/2}_{xy}$ is the in-plane effective electron mass discussed below, $\varepsilon_x$ and $\varepsilon_y$ are the in-plane excitation energies, $E_z$ is the electric field, and $x$, $y$, and $z$ are dot coordinates. In calculations, we take the limit $V_0 \to \infty$, so that the wave function is zero for $z \leq 0$
\footnote{The limit is subtle, as it does not commute with evaluating matrix elements of differential operators [App.~C in Ref.~\cite{carballido_low-symmetry_2019}, App.~B in \cite{simion_magnetic_2014}]. For crystal momentum expectation values, for powers higher than quadratic, it is necessary to account for artifacts of the hard-wall potential.
}. Having specified the confinement, we have
\begin{equation}
	H = H_{L} + \delta H_{L} + V_{xy}  +  V_{z},
	\label{eq:3Dqubit}
\end{equation}
as the full---three-dimensional---Hamiltonian describing the confined hole. 
We include the orbital effects of the in-plane magnetic field components via 
the vector potential, entering the momentum $\hbar \mathbf{k} = -i \hbar \nabla + e\mathbf{A}$. We fix the gauge to $\mathbf{A} = (z-z_0) (B_y,-B_x) + eB_z(-y,x) $, where $z_0$ is the ground state expectation value of the z coordinate.
Our next goal is to reduce this microscopic description into an effective Hamiltonian for the spin qubit, a two-level system. 

We first define the unperturbed Hamiltonian by supplementing the confinement by terms quadratic in momentum and not coupling the heavy hole and light hole subspaces,
\begin{equation}
	H_0^J =\frac{\hbar^2(\partial_x^2+\partial_y^2)}{2m^J_{xy}} + \frac{\hbar^2\partial_z^2}{2m^J_z} - V_{xy} -V_z.
	\label{eq:H0}
\end{equation}
The effective masses depend on the Luttinger parameters, the hole-spin subspace 
($J=|J_z|\in \{1/2, 3/2\}$ for the light and heavy hole, respectively),
and the growth direction. We give formulas for some cases of interest in Tab.~\ref{DotParameters} in the Supplementary Material. The unperturbed Hamiltonian defines the basis for the perturbation theory. 
Since it is separable in in-plane coordinates $x$ and $y$, the vertical coordinate $z$, and the spin, the 
basis states $| J, n_x, n_y, n_z\rangle$ can be indexed by four quantum numbers: the pair $(n_x, n_y)$ is the Fock-Darwin spectrum indexes, while $n_z$ labels eigenstates of the triangular potential, associated with energy scale $\epsilon_z^J=(\hbar e E_z/\surd{m_z^J})^{2/3}$ (see App.~A1 in Ref.~\cite{stano_orbital_2019} for details on the $z$-confinement eigenstates). The splitting of heavy and light holes $\Delta_\mathrm{HL}$ discussed in the introduction is the energy difference of the two ground states of Eq.~\eqref{eq:H0} for the two values of the spin index $J$. 

The qubit Hamiltonian follows by integrating out the orbital degrees of freedom, with the excited states taken into account within the second-order perturbation theory,
\begin{equation}
	\begin{aligned}
		\mathcal{H} = & \langle J, \mathbf{0} | \Big[ H  +   \sum_{(J^\prime, \mathbf{n})\neq(J,\mathbf{0})}  \frac{\delta H |J',\mathbf{n}\rangle \langle J',\mathbf{n}| \delta H  }{E_{|J,\mathbf{0}\rangle}-E_{|J',\mathbf{n}\rangle}}\Big] |J, \mathbf{0} \rangle.
	\end{aligned}
	\label{eq:perturbation}
\end{equation}
Here, $\delta H = H-H_0^J$, the summation is over all excited orbital states, the vector $\mathbf{n} = (n_x,n_y,n_z)$, and $E_{|J,\mathbf{n}\rangle}$ is the unperturbed eigenstate energy.  

This derivation follows the procedure of Refs.~\cite{stano_g_2018, stano_orbital_2019} with one difference. In those references, the reduction proceeded in two steps: first integrating out the vertical coordinate $z$, then the in-plane coordinates $x$ and $y$. 
Here, we do not neglect the in-plane excitation energies in the denominator of Eq.~\eqref{eq:perturbation}, as they are comparable to $\Delta_\mathrm{HL}$. On the other hand, we restrict the sum over $n_z$ in Eq.\eqref{eq:perturbation} to the lowest excited state [see Fig.~\ref{Fig:DeviceAndBandSchematic}(b)]. The restriction is suitable for quasi-two-dimensional dots, where out-of-plane excitation energies are larger than the in-plane ones. The resulting approximate form of $\mathcal{H}$ is the basis for the two main quantities of our work, the effective $g$-tensor, and the qubit energy. The dependence of the latter on the electric field is responsible for dephasing of the hole-spin qubit.

\begin{figure}[htp]
	\centering
	\includegraphics[trim=15 15 15 15,clip,width=0.40\textwidth]{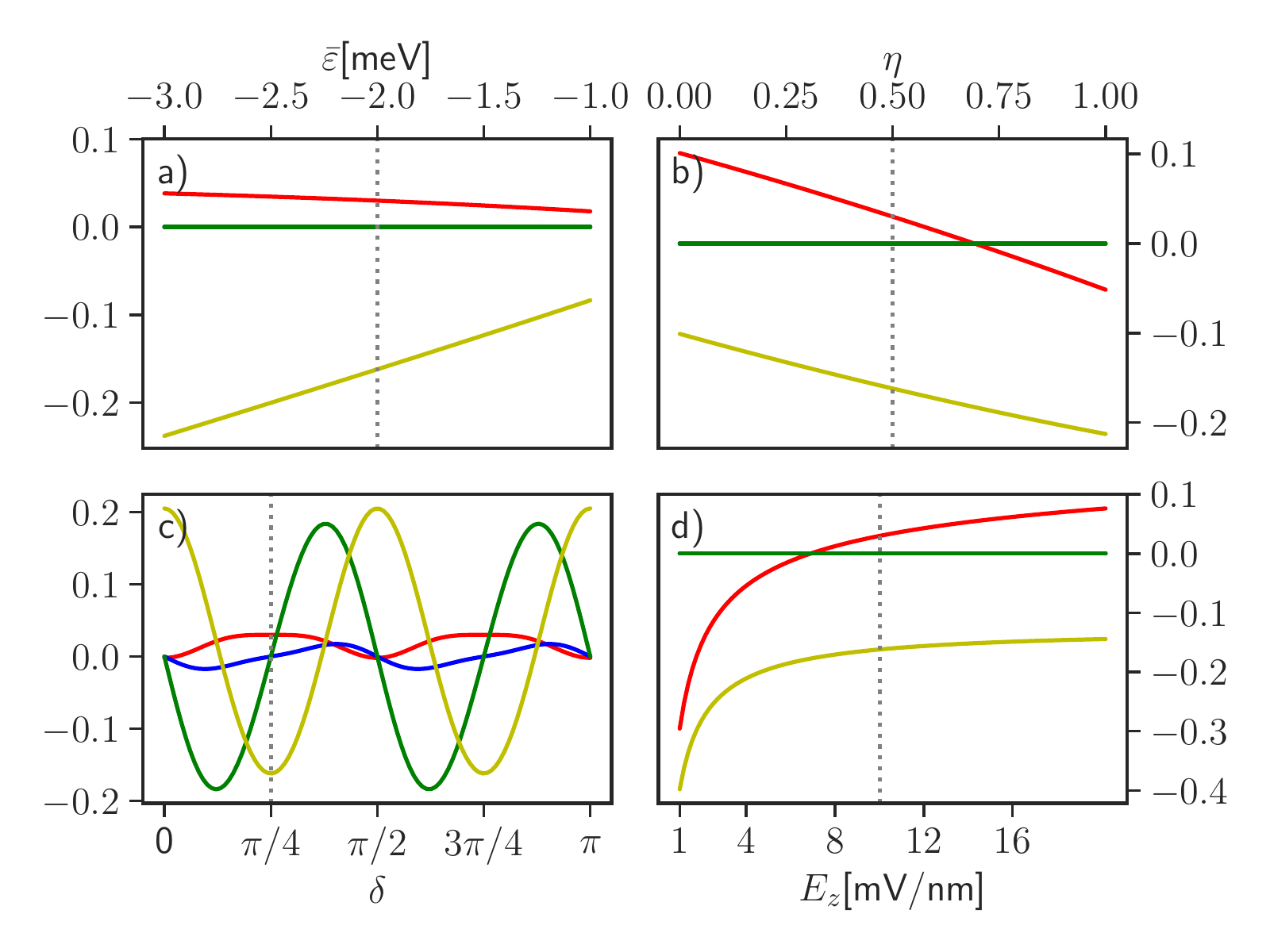}
	\caption{Heavy-hole qubit $g$-tensor components for a dot with $z_d$-axis along [001].	
		The colored lines show $\hat g_{xx}$(red), $\hat g_{xy}$(blue), $\hat g_{yx}$(green), and $\hat g_{yy}$(yellow). 
		The $x$-axis represents (a) the average dot energy $\bar{\varepsilon}= (\varepsilon_x+\varepsilon_y)/2$, (b) dot asymmetry $\eta =(\varepsilon_y-\varepsilon_x)/2\bar{\varepsilon}$, (c) dot orientation $\delta$, and (d) vertical confinement electric field $E_z$.
The dashed line, defined by $\varepsilon_x=-1 \text{ meV, } \varepsilon_y=-3 \text{ meV, } \delta = \pi/4 \text{ }$, and $E_z = 10 \text{ mV/nm}$, denotes a common reference point. }
	\label{fig:gfactorNoZ}
\end{figure}

{\it Effective $g$-tensor.}
Evaluating Eq.~\eqref{eq:perturbation} gives the Hamiltonian $\mathcal{H}$ describing the qubit as a two-level system. Due to the time-reversal symmetry, at zero magnetic field the two states are degenerate. Neglecting terms higher than linear in the magnetic field, an approximation that we adopt in evaluating Eq.~\eqref{eq:perturbation}, the latter becomes the Hamiltonian of a spin one-half,
\begin{equation}
	\mathcal{H}=\sum_{i,j=x,y,z} \mu_B  B_i  \hat g_{ij} \tau_{j}.
	\label{eq:gtensor}
\end{equation}
Here, $\boldsymbol{\tau}$ is a vector of Pauli matrices defined with up and down spin one-half states corresponding to perturbed 
spin states of $\mathcal{H}$, 
where the perturbation in Eq.~\eqref{eq:perturbation} mixes to the HH ($J_z=\pm 3/2$) spin ground state the LH ($J_z=\pm 1/2$) spin states,
and $\hat g$ is a second-rank tensor, the $g$-tensor. 
We have thus reduced the three-dimensional qubit-description of Eq.~\eqref{eq:3Dqubit} to a much simpler effective two-level  model. Nevertheless, this model reflects orbital effects through the $g$-tensor dependence on confinement electric fields, which we now examine.

Figure \ref{fig:gfactorNoZ} shows the $g$-tensor for a quantum dot with the $z$-axis along [001]. The $g$-tensor in-plane components are plotted as functions of the lateral dot size (panel a), its asymmetry (panel b), its orientation (panel c), and the vertical-confinement strength (panel d). The components off-diagonal in the in-plane versus vertical coordinate groups, $g_{xz}$, $g_{zx}$, $g_{yz}$, $g_{zy}$, are zero. 
The out-of-plane component $g_{zz}$ is typically an order of magnitude larger than the in-plane ones, and does not depend appreciably on any parameter except of the vertical electric field.
We include $g_{zz}$ in Fig.~\ref{fig:gfactorIncludingZ} of the Supplementary Material.
The $g$-tensor is strongly anisotropic, a consequence of the confinement breaking all symmetries of the crystal 
\cite{scappucci2020,gradl_asymmetric_2018}.
In realistic samples which are neither perfectly symmetric nor aligned with any particular direction with respect to the crystal axes one expects large variations of the $g$-tensor components. 
Most importantly, the $g$-tensor components clearly depend on the confinement electric field. Through this dependence, the charge noise influences the qubit energy and causes dephasing. We are now in a position to estimate the resulting dephasing time.

{\it Coherence time.} The charge noise in the sample and experiment electronics leads to fluctuations of the electric field at the dot location, and thereby to fluctuations of the qubit energy. The connection can be seen by writing the qubit energy, using Eq.~\eqref{eq:gtensor}, as
$\hbar \omega = \sqrt{\mu_B^2 B_jB_k g_{ji}g_{ki} }$, where repeated indices are summed over. The electric field enters the $g$-tensor in two ways: by defining the shape of the dot confinement and by inducing bandstructure terms. We find that the latter terms have negligible influence; this is the reason why no electric-field generated term is listed in Eq.~\eqref{eq:NewTerms}. We also neglect fluctuating electric in-plane fields, since a uniform field does not change the shape of a harmonic confinement. 
The noise in the $z$-component of the electric field remains, changing the qubit energy through changes in the vertical confinement strength.  The noise is described by its spectrum, 
$S(\omega)\!\! =\!\!\int_{-\infty}^{\infty}\!\!d\tau' e^{i \omega \tau'}\! \langle E_z(0)E_z(\tau') \rangle,
$ 
with the bracket standing for a statistical average.  
We further assume that   1/f noise is dominant and take $S(\omega) = A / |\omega|$. The levels of noise measured in silicon samples were recently reviewed in Ref.~\cite{kranz_exploiting_2020}.
However, the majority of reports give them as fluctuations of the dot energy, not the electric field. 
We take $A = 2800 \text{ V}^2\text{/m}^2$ as measured in Ref.~\cite{kuhlmann_charge_2013} for the electric field fluctuations, which would yield dot energy fluctuations within the range reported in Ref.~\cite{kranz_exploiting_2020}. 

Reference \cite{ithier_decoherence_2005} finds that 1/f noise causes a Gaussian decay with a pure-dephasing rate
\begin{equation}
1/T_2^* =| \partial_{E_{z}} \omega| \sqrt{ A \ln [\omega_h/2\pi \omega_l]},
\label{eq:dephasing}
\end{equation}
where $\omega_l$ and $\omega_h$ are the low and high frequency cut-offs, respectively.
\begin{figure}[tp]
	\centering
	\includegraphics[trim=10 70 0 10,clip,width=0.90\linewidth]{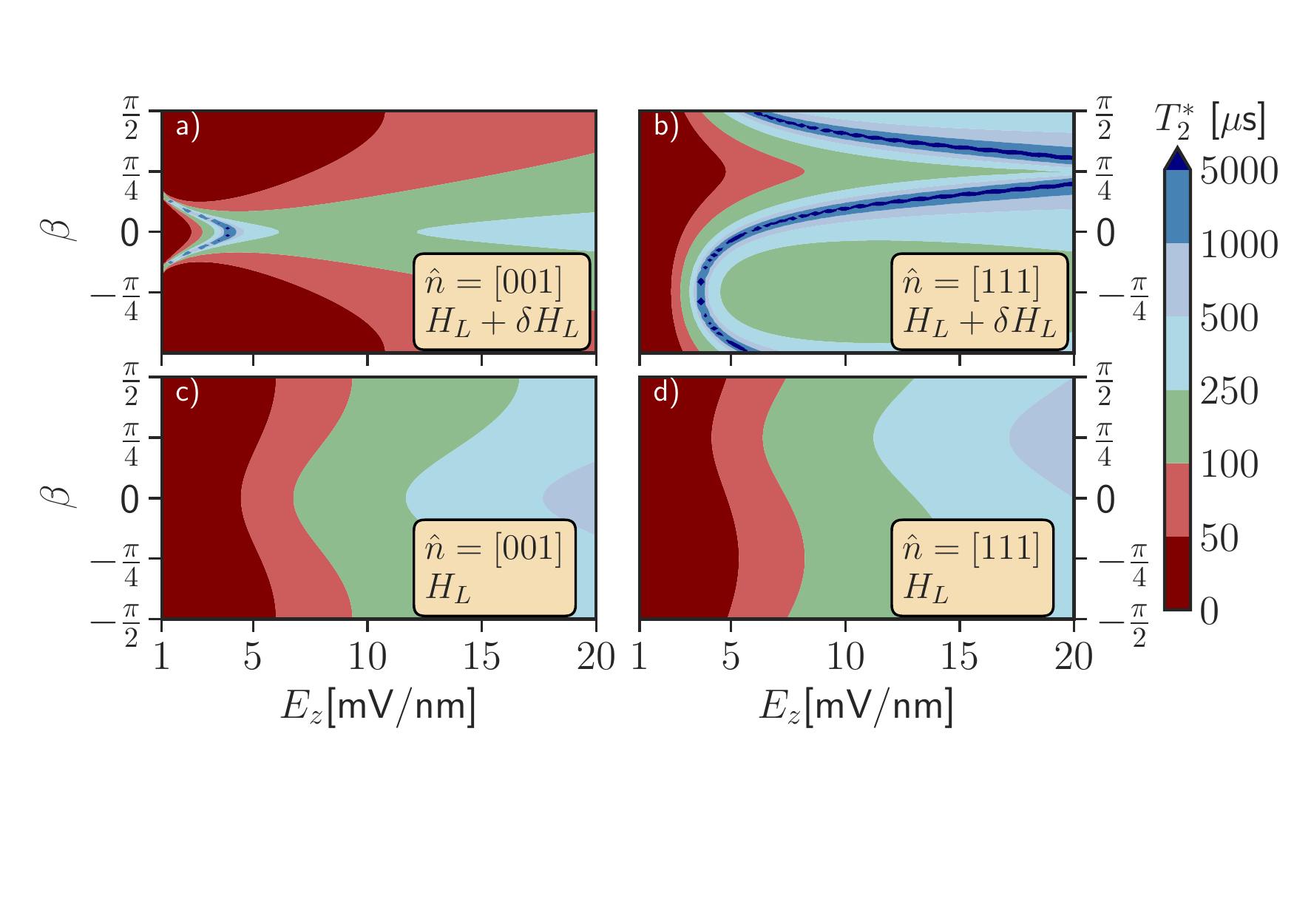}
	\caption{Dephasing time $T_2^*$ for spin-qubit devices grown along $\hat n =[001]$ and $\hat n=[111]$ when using the full Hamiltonian (top) and only $H_L$ (bottom). We consider a dot with $\delta=0$, an in-plane magnetic field $\mathbf{B}=B(\cos \beta, \sin \beta,0)$,
		where $B=1$ T, and $E_z$ is the vertical electric field. 
		We consider dot confinement lengths $l_x= \hbar ({m^{3/2}_{xy}}\varepsilon_x)^{-1/2} = 20$ nm, $l_y= 15$ nm, and
	$ \omega_h = 50$ kHz, $\omega_l=10^{-2}$ Hz for the frequency cut-offs. }
	\label{Fig:T2}
\end{figure}

Our main result is the analysis of the dephasing time $T_2^*$ calculated from Eq.~\eqref{eq:dephasing}. We plot the dephasing time in Fig.~\ref{Fig:T2}, varying the electric field strength $E_z$ and the in-plane magnetic field orientation for fixed strength $B=1$ T.
Figure \ref{Fig:T2}a (\ref{Fig:T2}c) shows the dephasing time $T_2^*$ for devices grown along [001] when including (excluding) the $\delta H_L$ terms. We note that in both models, the dephasing time is maximized for the $B$-field along $x$-axis, i.e. $\beta=0$, except at small electric fields in Fig.~\ref{Fig:T2}a. The dephasing time ranges from tens to hundreds of $\mu$s in most of the plot area. 
However, including $\delta H_L$ reveals a line of sweet spots with the dephasing time boosted up to milliseconds. 
For the parameter space considered in Fig.~\ref{Fig:T2}, the sweet spot line appears only within a limited range of the magnetic-field directions, $|\beta| \lesssim \pi/8$, and electric field magnitudes, $E_z \lesssim 5V/\mu$m.
	The appearance of sweet spots is not specific to the [001] growth direction. To illustrate this, Fig.~\ref{Fig:T2}b and Fig.~\ref{Fig:T2}d show the dephasing for [111]. 
	The situation is analogous, though there are quantitative differences: First, the discrepancy from not including the $\delta H_L$ terms is even more drastic. Second, the sweet spot appears within a larger range of the electric fields and for any orientation of the magnetic field. 
Since we found similar sweet spots for [011] growth direction (plots not shown), 
we conjecture that sweet spots in lateral spin hole qubits are generic \cite{wang_ultrafast_2020, bosco_hole_2020}.

Before concluding, we review available experimental results on the confined hole spin dephasing times. There are not many: An optical probe of a single hole in a III-V self-assembled dot gave $T_2^*$ of 100 ns in Ref.~\cite{brunner_coherent_2009} and 0.5 $\mu$s in Ref.~\cite{prechtel_decoupling_2016}, with the coherence time $T_2$ estimated an order of magnitude larger. 
The only numbers we are aware of in silicon is $T_2^*=60$ ns from Ref.~\cite{maurand_cmos_2016}, which was prolonged four-fold upon Hahn-echo, as expected for a 1/f noise \footnote{For 1/f noise, the dephasing time under a Hahn-echo equals \cite{ithier_decoherence_2005} $T_2^*$ times $\sqrt{\ln(\omega_h/2 \pi \omega_l)/\ln 2}$, evaluating to $4.4$ for the parameters given in the caption of Fig.~\ref{Fig:T2}.}, and $T_2^* = 440$ ns from Ref.~\cite{camenzind2021} for a hole spin qubit in a Si FinFET device.

While all these numbers are lower than our $T_2^*$, the difference is not so drastic considering that the charge noise levels have large variations among different materials and samples \cite{kranz_exploiting_2020, freeman_comparison_2016}. Coincidentally, the dephasing times that we obtained are comparable to nuclear-limited dephasing times of electrons in silicon dots: $T_2^*$ in natural silicon is a few microseconds, close to our values away from the sweet spot, and a millisecond in purified silicon-28, comparable to our sweet-spot values.
We conclude that the intrinsic material charge noise might be limiting coherence in some of these experiments, while in some, such as Si FinFETs, the nuclear noise is still the limiting factor, see Refs.\cite{camenzind2021,boscoprl2021}. 
Our results suggest that holes in planar dots can reach coherence comparable to electrons \footnote{Using the hyperfine coupling coefficient for silicon in Ref.~\cite{Philippopoulos2020} with Eq. 29 of Ref.~\cite{Struck2020}, we obtain $T_2^*=6\mu$s, while for purified silicon with 800ppm we get $T_2^*=0.3$ms.}, and searching for the hole-qubit sweet spots experimentally looks attractive.

\textit{Conclusions}.
In this work, we have quantified charge-noise induced dephasing rate of a silicon spin hole qubit, and the effective $g$-factor. For typical dot dimensions and external magnetic and electric fields, we find it is necessary to go beyond the Luttinger model to assess the $g$-tensor and dephasing reliably; the difference is qualitative. Our model, which can also be extended to devices using other diamond crystal materials, e.g., germanium hole spin devices, predicts a sweet spot for the dephasing time. We find that the sweet spots depend on the device growth direction, confinement potential, and in-plane magnetic field orientation. Our work leaves space for interesting extensions. For example, the dependence on the device geometry prompts the question of how the additional spin-orbit interactions impact spin dephasing in other devices, such as nanowire-based hole-spin qubits \cite{froning_2021} or FinFETs \cite{bosco_hole_2020}. 

This work was partially supported by the Swiss
National Science Foundation and NCCR SPIN.
\bibliographystyle{apsrev4-1}
\bibliography{_included}

\newpage

\begin{widetext}

\appendix
\section{Supplemental Material: ``Charge-noise induced dephasing in silicon hole-spin qubits"}

In the Supplemental Material we provide details used for our numerical calculations and justify our minimal model proposed for the spin dephasing in the main text.
In particular, we show in Table \ref{ParameterTable} the numerical coefficients that are used for the minimal model.
Moreover, we provide omitted results such as out-of-plane $g$-tensor components and dephasing times for devices with a dot-normal along [111]. For the latter result, we introduce the minimal set of terms, quartic in momentum, in Eq.~\eqref{Eq:NewTerms111}, that are needed to accurately reproduce the spin dephasing time $T_2^*$ shown in Fig. ~\ref{Fig:T2} of the main text.
In addition, we demonstrate the accuracy of our minimal model by comparing
its results with the ones obtained from the more general extended model, Eq.~\eqref{eq:3Dqubit},
used in the main text for both considered growth directions.

\begin{table}[htp]
	\begin{center}
		\begin{tabular}{|c|c|c|c|c|}
			\hline
			\multicolumn{2}{|c|}{Coupling strengths} & \multicolumn{2}{|c|}{ Band parameters  }                                            & Band structure \\
			\hline
			$\kappa_{41}$                            & \valueKappaFourOne$$ nm${}^{2}$                & $P$ & \valueBandP  eVnm     & \multirow{7}{*}{
			\includegraphics[width=0.15\textwidth]{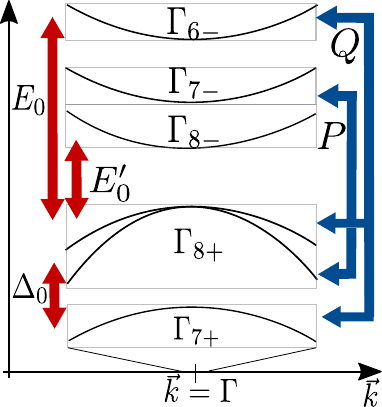}} 		  \\
			\cline{1-4}
			$ 	q_{41} $                               & \valueQFourOne$$ nm${}^{2}$                    & $Q$ & \valueBandQ   eVnm     & \\
			\cline{1-4}
			$\kappa_{42}$                            & \valueKappaFourTwo$$ nm${}^{2}$               & $E_0 $             & \valueEZero      eV   &   \\
			\cline{1-4}
			$ 	q_{42} $                               & \valueQFourTwo$$ nm${}^{2}$                  & $\Delta_0$         & \valueDeltaZero   eV  &   \\
			\cline{1-4}
			$ \kappa_{43}   $                        & \valueKappaFourThree$$ nm${}^{2}$              & $E_0'$             & \valueEZeroPrime  eV    & \\
				\cline{1-4}
			$			  q_{43}    $                           & \valueQFourThree$$ nm${}^{2}$                  & $\Delta_0'$        & \valueDeltaZeroPrime eV  &\\
			\cline{1-4}
			$			  \Gamma_{53}    $                      & \valueGammaFiveThree$$ nm${}^{2}$                   &                    &          &              \\
			\cline{1-4}
			$			  \Gamma_{32}    $                      & \GammaThreeTwo$$ nm${}^{4}$                   &                    &          &              \\
				\cline{1-4}

			$			  \Gamma_{12}    $                      & \GammaOneTwo$$ nm${}^{4}$                   &                    &          &              \\
			\hline
		\end{tabular}
		\caption{Prefactors (left box) as calculated in this work. The band parameters (middle box), defined by the schematic band structure (right box), are taken from Tables II and III of Ref. \cite{richard_energy-band_2004}.
			The energy splittings and the matrix elements coupling different bands around the high-symmetry point $\Gamma$ are shown by the red and blue lines, respectively.
		}
		\label{ParameterTable}
	\end{center}
\end{table}

The device growth orientation directly impacts the effective masses for holes occupying the heavy and light hole bands. Table~\ref{DotParameters} summarizes the properties of the states entering our perturbation terms.
\begin{table}[htp]
	\begin{center}
		\begin{tabular}{@{}c@{\quad}c@{\quad}c@{\quad}c@{\quad}c@{\quad}c@{\quad}@{}}
		\toprule
		 growth direction & hole character& total spin&  mass in the plane & mass along z\\
			&LH or HH & $J_z$           & $m_0/m_{xy}$ & $m_0/m_{z}$       \\
		\midrule
			\hline
		 $[001]$	& LH & $\pm 1/2$         &$\gamma_1 - \gamma_2$ & $\gamma_1 + 2\gamma_2$       \\
		 $[001]$	& HH & $\pm 3/2$         &$\gamma_1 + \gamma_2$ & $\gamma_1 - 2\gamma_2$       \\
		 $[111]$	& LH & $\pm 1/2$         &$\gamma_1 - \gamma_3$ & $\gamma_1 + 2\gamma_3$       \\
		 $[111]$	& HH & $\pm 3/2$         &$\gamma_1 + \gamma_3$ & $\gamma_1 - 2\gamma_3$       \\
		 \bottomrule
		\end{tabular}
		\caption{The effective masses for the HH and LH states in devices with [001] and [111] growth directions, which are needed in the numerical calculation.
				}
							\label{DotParameters}
	\end{center}
\end{table}
In the following we demonstrate the accuracy of the minimal model proposed in the main text, and show that an analogous model is available for devices with a dot-normal along the [111] crystallographic axis.
The analogous model comprises the following set of terms, and those contained within the Luttinger Hamiltonian $H_L$, and describes accurately the spin dephasing behaviour, as shown in Fig. \ref{Fig:T2MinimalModel}:
\begin{equation}
	\begin{aligned}
		H_{42}= & \mu_B(\kappa_{42} \mathbf{J} + q_{42}\mathbf{J}_3) \cdot  (B_X  k_X^2, B_Y  k_Y^2, B_Z  k_Z^2), \\
		H_{43}= & \mu_B(\kappa_{43} \mathbf{J} + q_{43}\mathbf{J}_3) \cdot  (k_X (k_Y B_Y + k_Z B_Z), \mathrm{c.p.}),                  \\
		H_{32}= & \Gamma_{32} \mathbf{J}_{32} \cdot (2k_X^2 k_Y^2 - (k_Y^2+ k_X^2)k_Z^2,  (k_Y^2  - k_X^2)k_Z^2).   \\
		H_{51} =& \Gamma_{51}(\{k_X^2,k_Y,k_Z\} J_{YZ} + \{k_Y^2,k_Z,k_X\} J_{ZX}+ \{k_Z^2,k_X,k_Y\}J_{XY}),\\
	\end{aligned}
	\label{Eq:NewTerms111}
\end{equation}
where $\{a,b,c\}=(abc + acb + cba + bca +bac+bca)/6$ is used and the prefactor $\Gamma_{51}=\valueGammaFiveOne$ nm${}^{4}$ is introduced.
\begin{figure}[tp]
	\centering
	\includegraphics[width=0.5\linewidth]{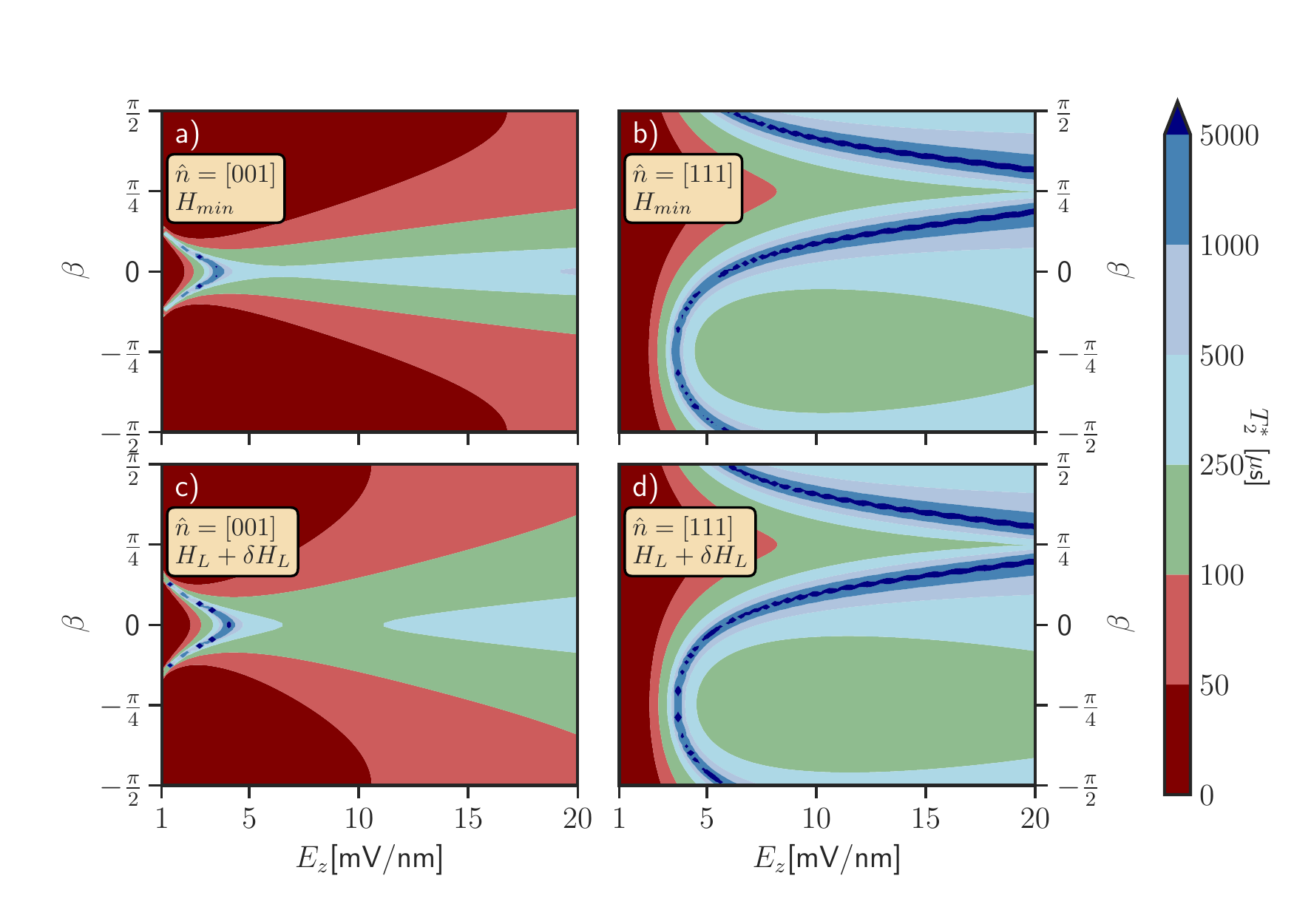}
	\caption{Estimated dephasing time $T_2^*$ (color coded) for a single hole spin qubit in a planar quantum dot grown along $\hat n =[001]$ and $\hat n=[111]$ when using the minimal Hamiltonian $H_{min}$ (top) and full Hamiltonian $H_L+\delta H_L$ (bottom).  The minimal Hamiltonian $H_{min}$ contains only the subset of terms described in the main text for [001]-devices and Eq. \eqref{Eq:NewTerms111} for [111]-devices.
		The dot axes are aligned with the crystallographic axes, the magnetic field is in-plane with the magnitude $B=1$T and direction $\beta$, and the vertical electric field is $E_z$. The same parameters for the dot shape and charge noise are used as in Fig.~\ref{Fig:T2}.
}
	\label{Fig:T2MinimalModel}
\end{figure}
In the main text we briefly mention the difference between the in-plane and out-of-plane components of the $g$-tensor. In Fig.~\ref{fig:gfactorIncludingZ} we show that the out-of-plane components are in general much larger than the in-plane ones and do not change significantly when varying the dot parameters.

\begin{figure}[htp]
	\centering
	\includegraphics[trim=15 10 15 10,clip,width=0.45\textwidth]{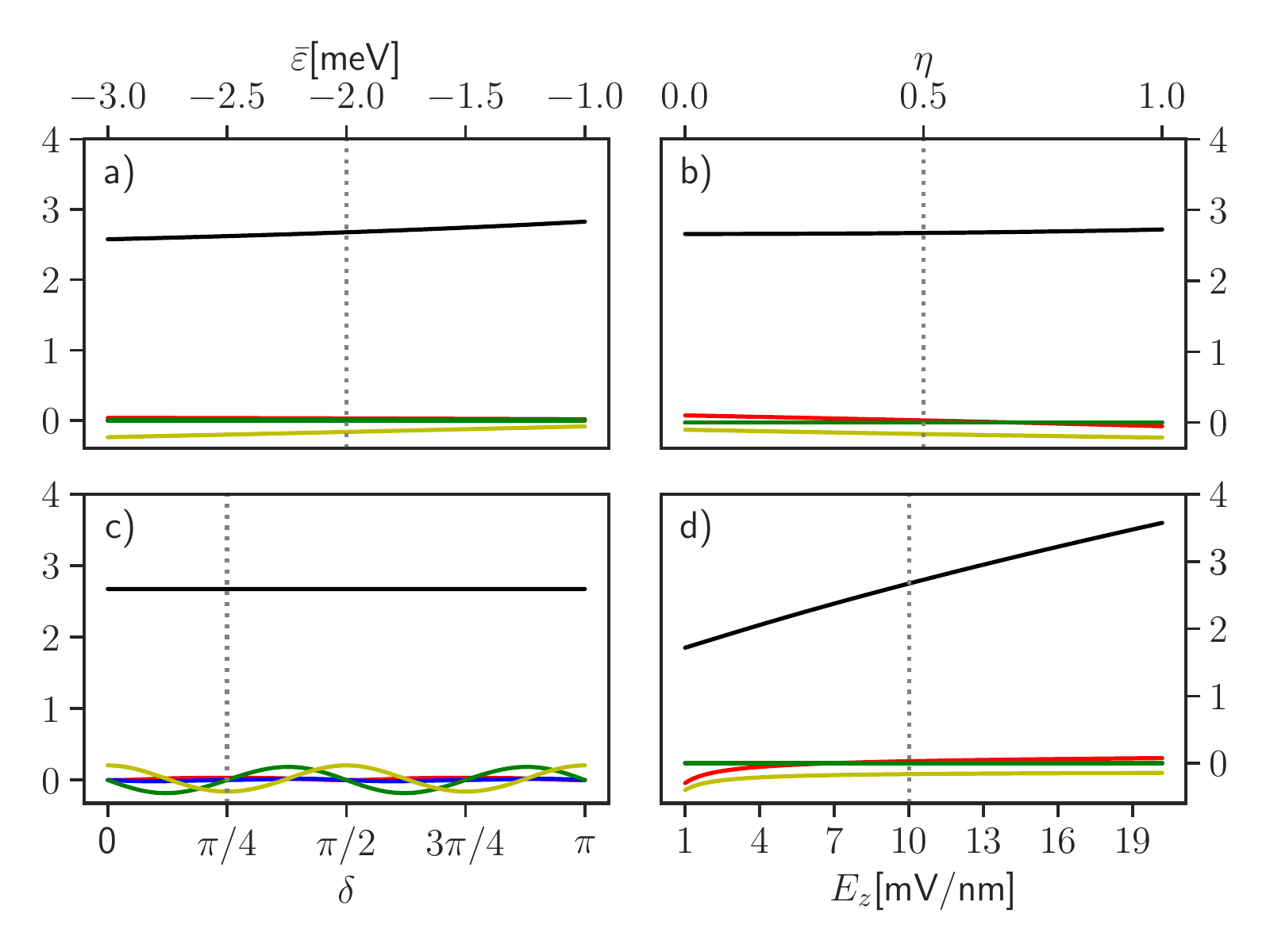}
	\caption{Components of the heavy-hole qubit $g$-tensor.
		The colored lines show $\hat g_{xx}$(red), $\hat g_{xy}$(blue), $\hat g_{yx}$(green), $\hat g_{yy}$ (yellow) and $\hat g_{zz}$ (black).
		The $x$-axis represents (a) the dot size $\bar{\varepsilon}= (\varepsilon_x+\varepsilon_y)/2$, (b) the dot asymmetry $\eta =(\varepsilon_y-\varepsilon_x)/2\bar{\varepsilon}$, (c) the dot orientation $\delta$, and (d)  the vertical confinement electric field  $E_z$.
The dashed vertical line in each panel, defined by $\varepsilon_x=-1 \text{ meV, } \varepsilon_y=-3 \text{ meV, } \delta = \pi/4 \text{ }$, and $E_z = 10 \text{ mV/nm}$, denotes a common reference point. }
	\label{fig:gfactorIncludingZ}
\end{figure}

\end{widetext}

\end{document}